
\magnification 1200
\font\abs=cmr9

\def\uno{{\bf 1}}
\def\tens{\otimes}
\def\fraz#1#2{{\strut\displaystyle #1\over\displaystyle #2}}

\def\ket#1{|~#1~\rangle}

\def\ii#1{\item{$[#1]~$}}

\def\mub{\mu_{\scriptscriptstyle B}}

\hsize= 15 truecm
\vsize= 22 truecm
\hoffset= 0.5 truecm
\voffset= 0 truecm

\baselineskip= 21 pt
\footline={\hss\tenrm\folio\hss} \pageno=1
\centerline
{\bf  QUANTUM GALILEI GROUP }
\smallskip
\centerline
{\bf AS SYMMETRY OF MAGNONS.}

\bigskip\bigskip
\centerline{
{\it    F.Bonechi ${}^1$, E.Celeghini ${}^1$, R. Giachetti ${}^2$,
             E. Sorace ${}^1$ and M.Tarlini ${}^1$.}}
\bigskip

   ${}^1$Dipartimento di Fisica, Universit\`a di Firenze and INFN--Firenze,
\footnote{}{\hskip -.85truecm E--mail: TARLINI@FI.INFN.IT\hfill
   DFF 156/03/92 Firenze}

   ${}^2$Dipartimento di Matematica, Universit\`a di Bologna and INFN--Firenze.
\bigskip
\bigskip

{\bf Abstract.} {\abs Inhomogeneous quantum groups are shown to be an
effective algebraic tool in the study of integrable systems and to provide
solutions equivalent to the Bethe ansatz. The method is illustrated on the
1D Heisenberg ferromagnet whose symmetry is shown to be the quantum Galilei
group $\Gamma_q(1)$ here introduced. Both the single magnon and the
$s=1/2$ bound states of $n$--magnons are completely described by the algebra.}
\smallskip
\noindent PACS 03.65.Fd; 75.10.Jm\hfill\break

\bigskip
\bigskip

In very recent times it has been observed that physical systems with a
fundamental length scale have a symmetry of a inhomogeneous quantum group [1].
Fruitful results have been obtained in nuclear physics, where the quantum
parameter is related to the time scale of strong interactions [2], as well
as in solid state physics, where a fundamental length arises naturally [3].
In this letter we show that, actually, the quantum
group symmetry yields an algebraic scheme consistent with the Bethe
ansatz [4,5] for solving the dynamics of quantum integrable models. We shall
illustrate the method on the concrete example of the Heisenberg ferromagnet.
By looking at the single magnon excitations, the symmetry of the model
is found to be a $q$-deformation of the Galilei group in one
dimension, where the deformation parameter is the lattice spacing. The
coalgebra structure will then define naturally the many magnon systems,
both producing the global variables and describing
bound states for $s=1/2$ as kinematical properties of the structure.

The Hamiltonian of the Heisenberg ferromagnet
with periodic conditions ${\vec S}_{N+1}={\vec S}_{1}$ is given by [6]:
$$ {\cal H} = - J \sum_{i=1}^N {\vec S}_i\cdot {\vec S}_{i+1} +
g \mub H\sum_{i=1}^N S^z_i\ , \eqno(1)$$
where ${\vec S}_i$ are spin $s$ operators,
$H$ is an external homogeneous magnetic field and the exchange integral $J$
is positive.
The energy of the ground state, reads $E_0 = - J N s^2 - g\mub H N s\ $
and the states with one spin deviate are
$\psi=\sum\limits_i f_i\; S^{+}_i\ket{0}$
with the usual definition of $S^{\pm}_i$.
The eigenvalue equation ${\cal H}\;\psi= E\;\psi$ is equivalent to the
algebraic system
$$-sJ(f_{i-1}+f_{i+1}-2f_{i})+g\mub Hf_{i} = (E-E_0)f_{i}\ .\eqno(2)$$
The diagonalization of equation (2) leads to the following dispersion relation
for plane waves with wave vector $k$:
$$E-E_0=2sJ(1-\cos k)+g\mub H\ .\eqno(3)$$

Let us now show that a quantum group symmetry can give an algebraic description
of the system.
Adopting the same point of view of ref. [3], we see that the solutions of
system (2) are obtained by evaluating at integer multiples of the lattice
spacing $a$ the solutions of the differential equation
$$\Bigl(2sJ(1-\cos(-ia\partial_x))+g\mub H\Bigr)f(x)=(E-E_0)f(x)\ .\eqno(4)$$
In the limit $a\rightarrow 0$, (4) is just the
Schr\"odinger equation in a constant potential, with an effective
mass $(2sJa^2)^{-1}$ and with the symmetry of the 1-dim Galilei group.

We introduce a new inhomogeneous quantum algebra, that we call
$\Gamma_q(1)$, which is a deformation of the Galilei algebra and
is generated by the four elements $B,\ M,\ P,\ T.\ $ The commutation
relations are:
$$ [B,P]=iM\ ,~~~~~~~~~~[B,T]=(i/a)\sin(aP)\ ,~~~~~~~~~~[P,T]=0\ ,$$
the generator $M$ being central.

The coproducts read
$$ \eqalign{
 \Delta B=e^{-iaP}\tens B+B\tens e^{iaP}\,,~~~~~~&~~~~~~
     \Delta M=e^{-iaP}\tens M+M\tens e^{iaP}\,,\cr
 \Delta P = \uno\tens P+P\tens\uno\,,~~~~~~&~~~~~~
 \Delta T = \uno\tens T+T\tens\uno\,,\cr}$$
and the antipodes
$$\gamma(B)=-B-aM\,,~~~~~\gamma(M)=-M\,,~~~~~\gamma(P)=-P\,,~~~~~
\gamma(T)=-T\,.$$
The Casimir of $\Gamma_q(1)$ is
$$C=MT-(1/a^2)\;(1-\cos(aP))\,.$$

A differential realization of this quantum algebra is given by
$$\eqalign{
B&=mx\,,~~~~~~~~~~M =m\,,~~~~~~~~~~P=-i\partial_x\,,\cr
T&=(ma^2)^{-1}(1-\cos(-ia\partial_x))+c/m\,,\cr}\eqno(5)$$
where $c$ is the constant value of the Casimir $C$:
for $m=(2sJa^2)^{-1}$ and $c=mg\mub H$ we see that
the expression of $T$ coincides with the left hand side of (4).

The algebra is invariant under $P\mapsto P + (2\pi/a)\,n$, so that we can
choose for $P$ any domain of size $2\pi/a$. Like in the Galilei algebra the
position operator is defined as $X=(1/M)\, B$, with time derivative
$$\dot X\;=\;i[T,X]\;= \;2 s J a\; \sin (a P)\,, $$
in agreement with the canonical equation.
The properties concerning the single magnon states are then
obtained using the symmetry under the quantum algebra $\Gamma_q(1)$.

We are now going to study a system of two magnons in a Heisenberg ferromagnet
described by the state $\psi=\sum_{i>j} f_{ij}\;
S^{+}_iS^{+}_j\ket{0}$, with $f_{ij}=f_{ji}$.
The eigenvalue equation for the Hamiltonian leads to the following system
for $f_{ij}$:
$$\eqalign{
(E - E_0 - &2g\mub H - 4sJ)\;f_{ij} +
s\sum_n (J_{nj}\,f_{in} + J_{in}\,f_{nj}) \cr
{}&=\fraz12 J_{ij}(f_{ii} + f_{jj} - f_{ij} - f_{ji})\ ,\cr} \eqno(6)$$
where the bonds $J_{ij}$ are equal to $J$ when the label $(ij)$ are nearest
neighbor pairs
and vanish otherwise. For $s=1/2$ the amplitudes $f_{ii}$ are unphysical:
however such amplitudes appear in both sides of (6) for $i=j+1$ and cancel.
The left hand side of (6) constitutes the equation for the free part of the
system, while the terms of the right hand side are responsible for the
interaction. The quantum group
$\Gamma_q(1)$ is a symmetry of the free system: nevertheless it is able to
describe many magnon states, provided that the interaction can be treated as a
``boundary condition'' ensuring that the homogeneous free equation is satisfied
at every pair of sites. This is indeed the Bethe ansatz, which imposes
the separate vanishing of the two sides of equation (6).

The coproduct describes the composition of
elementary systems. When the symmetry is given by a Lie algebra,
since the generators are primitive, the global symmetry of a composite
system is obtained by summing the generators of the algebra of the
elementary constituents.
In the quantum group context we can have non primitive generators, but
the coproducts can be used for the same purpose.

{}From the coproduct of $T$,  $\Delta T = \uno\tens T+T\tens\uno$, we
find for the energy $T_{12}$ of a two magnon system
$$\eqalign{
T_{12}&=T_1 + T_2 \cr
      {}&= (M_1 a^2)^{-1} (1-\cos(aP_1)) +
  (M_2 a^2)^{-1}(1-\cos(aP_2))  + C_1/M_1 + C_2/M_2\ .\cr}$$
Considering magnons with the same value of $s$, the eigenvalue of $M_1=M_2=M$
is equal to $(2sJa^2)^{-1}$ and $C_2/M_2=C_1/M_1=g\mub H$.
Using the differential
realization $P_1=-i\partial_{x_1}$ and $P_2=-i\partial_{x_2}$ we get
$$\eqalign{
T_{12}\,f(x_1,x_2)=&\,4sJ\,f(x_1,x_2) + 2g\mub H\,f(x_1,x_2) -
                                     sJ\Bigl(f(x_1,x_2+a)\cr
              {}& + f(x_1,x_2-a) + f(x_1+a,x_2) + f(x_1-a,x_2)\Bigr)\ \cr}
\eqno(7)$$
and the eigenvalue equation for $T_{12}$ coincides with the vanishing of the
left hand side of equation (6). Plane waves solve the equation (7) with energy
$$E-E_0 = 2sJ\Bigl(2 - \cos(ap_1) - \cos(ap_2)\Bigr) + 2g\mub H\ .$$
This is the energy of the continuum, the eigenfunction of the two magnons
states being obtained by imposing the periodic boundary conditions on the
plane waves.

An important point is that the interaction coefficients $J_{ij}$ depend
on the relative coordinates, so that the total momentum is always a
constant
of the motion. This is rephrased in our algebraic approach by the fact that
$P$ is a primitive generator, {\it i.e.} $\Delta P = \uno\tens P+P\tens\uno$;
the quantum group implies then
$$P_{12}=P_1+P_2\ ,$$
and the total momentum  has the correct symmetry and invariance in the
composite system.

For $s=1/2$ a solution for the bound states based on the
Bethe ansatz is known [4].
We show here that our treatment based on $\Gamma_q(1)$
reproduces this result in a very natural way. For higher spin neither
the Bethe ansatz nor our quantum group approach apply, since the model
does not reduce to a free equation plus ``boundary conditions".

The generator $M$ is a central element in the algebra, but the coalgebra
shows that it combines in a nontrivial way:
$$M_{12}=M_1 e^{iaP_2} + M_2 e^{-iaP_1}\ .$$
We then rewrite $T_{12}$ in terms of global and ``relative energy"
$$T_{12}= (a^2\,M_{12})^{-1} (1-\cos(aP_{12})) +
C_1/M_1 + C_2/M_2 - \fraz{(M_{12} - M_1 - M_2)^2}{2\,a^2\,M_{12} M_1M_2}\ ,
\eqno(8)$$
where the ``relative energy" is invariant with respect to the global variables
and in the continuum limit $\ a\rightarrow 0\ $ gives the Galilei
relative energy $\ \fraz{M_1M_2}{2(M_1+M_2)} (P_1/M_1 - P_2/M_2)^2\ .$
The only new feature with respect to the Lie case is that the effective mass
of the composite system is a complex variable.
For $s=1/2$ magnons we have $M_1=M_2=M=(Ja^2)^{-1}$.

In (8) the relative energy is vanishing on the states for which
$M_{12}=2M$, namely
$$M_{12}\,f(x_1,x_2)=M\,\bigl(f(x_1,x_2+a)+f(x_1-a,x_2)\bigr)=
2M\,f(x_1,x_2)\ .\eqno(9)$$
Searching a solution of the form
$$f(x_1,x_2)=e^{i p (x_1+x_2)/2} F(x_1-x_2),\quad\quad p=p_1+p_2\ ,
\quad\quad x_1>x_2\ ,$$
{\it i.e.} solving the problem at fixed $P_{12}$ and using the invariance
of $p$ under translations of $2\pi n/a$,
equation (9) for $F(x_1-x_2)$ takes the form
$ F(x_1-x_2) = |\cos(ap/2)|\;F(x_1-x_2-a)\ $ and is solved by
$$F(x_1 - x_2)=const\cdot\left|\cos(ap/2)\right|^{(x_1-x_2)/a}\ .$$
This solution is consistent with periodic
conditions only for asymptotically large values of
$N$; the evaluation of the energy on $f(x_1,x_2)$ results in
$$T_{12}\,f(x_1,x_2)=\Bigl((J/2)\bigl(1-\cos(ap)\bigr)+2g\mub H\Bigr)
                       \,f(x_1,x_2)\ .$$
This is the energy and the eigenfunction found by Bethe [4]. Indeed the
Bethe equation for bound states is nothing else than the constraint on the
coproduct, $M_{12}=2M$:
$$\cot(a P_1/2)-\cot(a P_2/2)=-2\,i\ \quad \Leftrightarrow \quad
e^{-iaP_1} + e^{iaP_2}=2\ \eqno(10)$$

It seems therefore clear that these bound states are not
a consequence of an actual interaction: their existence is due to the
lattice and they merge in the continuum for $a\rightarrow 0$.

Let us now consider the position operator of the composite system; from the
coproduct of $B$ we can define
$$\eqalign{
X_{12}=\fraz{B_{12}}{M_{12}}&=\fraz{x_2\,e^{-iaP_1}+x_1\,e^{iaP_2}}
           {e^{-iaP_1} + e^{iaP_2}}\cr
     {}&=\fraz{x_1+x_2}2 + i\,\fraz{x_1-x_2}2\;\tan(\fraz{aP_{12}}2)\ ,\cr}$$
which satisfies $[X_{12},P_{12}]=i\ $. For the bound state we find
$$\eqalign{
{\dot X}_{12}=i[T_{12},X_{12}]&=i[(J/2)(1-\cos(aP_{12}))+
                 2g\mub H,\;X_{12}]\cr
                    {}&=(J/2)\;a\sin(aP_{12})\ ,\cr}\eqno(11)$$
showing the same structure as for the one magnon states with twice the
value of the effective mass.
For continuous spectra of $P_1$ and $P_2$, {\it i.e.} $N\rightarrow \infty$,
the eigenvalue equation for $X_{12}$ can be defined in the momentum space:
$$X_{12}\,f(p,k)=\Bigl(i\partial_p - (1/2)\tan(ap/2)\,\partial_k\Bigr)\,f(p,k)=
\lambda\,f(p,k)\ ,\eqno(12)$$
where $p=p_1+p_2$ and $k=(p_1-p_2)/2$. The characteristic curves of equation
(12) are
$$k+(i/a)\,\ln|\cos(ap/2)|=const\ .$$
Thus, for hermitian $X_{12}$ and $P_{12}$, $k$ is imaginary and compatible
with a constant $\Delta M$, and thus with (10).

Let us now consider the generalization to the $n$--magnon case. Here also
the quantum group yields the relations giving the total energy
and the energy of the bound states. An induction procedure gives
$$\eqalign{
T_{12\dots n}=\sum_{k=1}^n T_k&=(a^2 M_{12\dots n})^{-1}
\Bigl(1-\cos(aP_{12\dots n})\Bigr) + \sum_{k=1}^n (C_k/M_k)\cr
  {}&-\fraz 1{2a^2}~\sum_{k=2}^n \fraz{(M_{12\dots k} - M_{12\dots (k-1)}
- M_k)^2}{M_{12\dots k} M_{12\dots (k-1)}M_k}\ }\eqno(13)$$
where $P_{12\dots n}=\sum_{k=1}^n P_k$ and $M_{12\dots n}$ is defined by
iterating the coproduct and using the coassociativity:
$$M_{l\dots k} = M_{l\dots (h-1)}\ e^{ia(P_h+\dots+P_k)} +
M_{h\dots k}\ e^{-ia(P_l+\dots+P_{h-1})}\ ,\quad\quad l< h\leq k\ .$$

In analogy with the two magnon case, the bound
states are obtained by imposing the vanishing of the relative energy in (13):
$$E^B_{12\dots n}=(J/n)\Bigl(1-\cos(ap)\Bigr)+n\,g\mub H\ ,\quad\quad
p=\sum_{k=1}^n p_k\ .$$
The conditions we get, namely $\ M_{1\dots k}=kM\ $ for $k=2,\dots,n$
are equivalent to the Bethe conditions $\ M_{(k-1)k}=2M $.

We conclude by giving some remarks.

The quantum group
$\Gamma_q(1)$ is applicable to any elementary physical system described by a
discretized Schr\"odinger equation; moreover it can be used in the same
way as the Galilei group in the continuum to account for the asymptotic
states in the presence of well behaved potential.

A Galilei continuous symmetry is present in the Bose gas model where the
Bethe ansatz consists in boundary conditions in the first derivatives [5,7];
we notice that our approach contains the description of this case in the
limit $a\rightarrow 0$ once we modify the coproduct constraint in
$M_{12}=2 M + a\; {\cal C} M$, where ${\cal C}$ is the delta potential
strength.

It is then interesting
to realize that it is just the appropriate quantum group symmetry that
indicates the Bethe ansatz and then the integrability of the system.

\bigskip
\bigskip
\noindent{\bf Acknowledgments.}
The authors thank A.G. Izergin and V. Tognetti for a critical reading of the
manuscript and for useful discussions.

\bigskip
\bigskip
\bigskip

\centerline{{\bf References.}}

\bigskip
\ii 1 E. Celeghini, R. Giachetti, E. Sorace and M. Tarlini, J. Math. Phys.
      {\bf 31}, 2548 (1990); J. Math. Phys. {\bf 32}, 1155 (1991);
      J. Math. Phys. {\bf 32}, 1159 (1991);\hfil\break
      E. Celeghini, R. Giachetti, E. Sorace and M. Tarlini,
      ``{\it Contractions of quantum groups}'', Proceedings of
      the first semester on quantum groups, Eds. L.D. Faddeev
      and P.P. Kulish, Leningrad October 1990, Springer-Verlag, in press.
\smallskip
\ii 2 E. Celeghini, R. Giachetti, E. Sorace and M. Tarlini,
        ``{\it Quantum Groups of Motion and Rotational Spectra of Heavy
               Nuclei.}'', Phys. Lett. B (1992), in press.
\smallskip
\ii 3 F. Bonechi, E. Celeghini, R. Giachetti, E. Sorace and M. Tarlini,
        ``{\it Inhomogeneous Quantum Groups as Symmetry of Phonons.}'',
        University of Florence Preprint, DFF 152/12/91.
\smallskip
\ii 4 H.Bethe, Z.Phys. {\bf 71}, 205 (1931).
\smallskip
\ii 5 V.E. Korepin, A.G. Izergin and N.M. Bogoliubov, ``{\it Quantum Inverse
Scattering Method and Correlation Function. Algebraic Bethe Ansatz}",
(Cambridge, to appear in 1992).
\smallskip
\ii 6 D.C. Mattis, ``{\it The Theory of Magnetism, I }", (Springer--Verlag,
      Berlin Heidelberg 1981).
\smallskip
\ii 7 E.H. Lieb and W. Liniger, Phys. Rev. {\bf 130}, 1605 (1963).

\bye